\documentclass[12p]{article}
 \usepackage{dcolumn}

\usepackage{latexsym}
\usepackage{cite}
\usepackage{color}
\usepackage{amsmath}
\usepackage{epsfig}
\usepackage{hyperref}
\newcommand{\ortala}[1]{\begin{center}#1\end{center}}

\newcommand{\sandd}[1]{\left\langle #1\right\rangle}

\newcommand{\summ}[3]{{{\underset{#1 }{\overset{#2}{\displaystyle\sum}}}#3}}
\newcommand{\prodd}[3]{{{\underset{#1
}{\overset{#2}{\displaystyle\prod}}}#3}}

\newcommand{\re}[1]{(\ref{#1})}

\newcommand{\eq}[2]{\begin{equation}\label{#1}  #2\end{equation}}

\newcommand{\paran}[1]{\left(#1\right)}

\newcommand{\sch}[1]{Schrodinger}
\newcommand{\komb}[2]{\paran{\begin{array}{c} #1 \\ #2 \end{array}}}

\linethickness{0.6pt}

\begin{document}

\ortala{\textbf{On the hysteresis behaviors of the higher spin Ising model}}

\ortala{\"Umit Ak\i nc\i\footnote{umit.akinci@deu.edu.tr}}
\ortala{\textit{Department of Physics, Dokuz Eyl\"ul University,
TR-35160 Izmir, Turkey}}

\section{Abstract}\label{abstract}

Hysteresis characteristics of the general Spin-S ($S>1$) Blume-Capel model have 
been studied within the effective field
approximation. Particular emphasis has been paid on the large negative valued crystal field region 
and it has been demonstrated for this region that, Spin-S Blume-Capel model has  
$2S$ windowed hysteresis loop in low temperatures. Some interesting results 
have been obtained such as nested characteristics of the hysteresis loops of successive 
spin-S Blume-Capel model. Effect of the rising crystal field and temperature on 
these hysteresis behaviors have been investigated in detail and physical mechanisms have been given.

\section{Introduction}\label{introduction}

Higher spin Ising model ($S>1$) is very important for understanding of the real 
magnetic materials. Althought $S-1/2$ problems are the most widely studied in the 
literature on the theoretical side, it is a well known fact that $S-1/2$  systems 
are highly idealized systems. For instance none of the known 
ferromagnetic/antiferromagnetic atom in the periodic table has $1/2$. When the 
atoms brought together in a solid, different spin values emerge due to the 
overlapping of the atomic orbitals of the constituent atoms. Indeed, there are 
numerous molecules that have very high spins in the ground state, for instance 
$S-6$\cite{ref1}, $S-8$, $S-10$ \cite{ref2}. Besides, the most of the magnetic 
materials are represented by higher spin systems. For instance bimetalic Prussian 
blue analogs $CsNi^{II}[Cr^{III}(CN)_6].(H_2O)$ can be represented by $S-3/2$, 
$S-1$, $S-2$ atoms on the lattice \cite{ref3}. Altought it is very important to 
investigate higher spin systems, there is a downward trend with the rising spin 
value in the theoretical literature. This is due to the fact that, rising 
computational time for simulations and rising mathematical difficulties for 
approximation schemes for greater spin values.

Ising model including the crystal field or the single-ion anisotropy was 
introduced as a
$S-1$ Blume - Capel (BC) model \cite{ref4,ref5}. Later on, it was generalized to 
the higher spin problems and solved within the mean field approximation (MFA) 
\cite{ref6}. Some variants of the model exist such as spin-S model with 
biquadrartic exchange interaction and it was solved within the cluster variation 
method \cite{ref7,ref8}. Also, transverse Ising model with higher spin has been 
solved within the effective field theory (EFT) \cite{ref9,ref10,ref11}. Quenched 
disorder effects such as site dilution has been also investigated for spin - S 
BC model with EFT \cite{ref12,ref13} and random crystal field problem within the 
pair approximation \cite{ref14}. Some other techniques such as Monte Carlo (MC) 
simulation exist for the  higher spin BC model \cite{ref15}. This short 
literature was for the general spin-$S$ models. If we look at the specific spin 
valued BC model, we see downward trend with rising spin values.

$S-3/2$ BC model have been introduced earlier to explain the phase transitions 
in $DyVO_4$  \cite{ref16,ref17} and tricritical properties of the ternary fluid 
mixtures  \cite{ref18}. This model has been solved by various methods such as EFT 
\cite{ref19,ref20,ref21,ref22}, two spin cluster approximation \cite{ref23}, 
thermodynamically self consistent 
theory \cite{ref24}, finite size scaling \cite{ref25}, MC and density matrix 
renormalization group technique   \cite{ref26}, renormalization group technique 
(RG) \cite{ref27}. Antiferromagnetic BC model has already been worked within the 
MFA \cite{ref28}, cluster variation method \cite{ref29}, transfer matrix 
technique \cite{ref30},  
transfer-matrix finite-size-scaling calculations and MC \cite{ref31}. Besides, 
$S-3/2$ BC model with bilinear and  biquadratic interactions has been 
investigated \cite{ref32,ref33,ref34}.  If we look at different geometries we 
can see that $S-3/2$ model within different geometries have also been worked 
such as Bethe lattice \cite{ref35,ref36,ref37},  multilayer \cite{ref38,ref39},
cylindrical Ising nanotube \cite{ref40}, hexagonal Ising nanowire \cite{ref41} 
and semi infinite geometry  \cite{ref42}.  Some generalizations of the $S-3/2$ 
Ising model have already been constructed and solved.  $S-3/2$ Ising model with 
transverse field 
\cite{ref43,ref44,ref45,ref46,ref47,ref48,ref49,ref50,ref51}, and transverse crystal 
field \cite{ref52,ref53} are among them. Quenched disorder effects, which is 
another important topic is widely worked out for the $S-3/2$ Ising model.   
Effect of the discrete random longitudinal field distribution (bimodal 
distribution) \cite{ref54,ref55,ref56,ref57,ref58} and random crystal field 
distribution \cite{ref59,ref60,ref61}, random bond distribution 
\cite{ref62,ref63,ref64}, bond dilution \cite{ref65}, site dilution 
\cite{ref66,ref67} problems have been worked out. Note that, EFT has been used 
in most of these quenched disorder works.

$S-2$ Ising model is also important due to the experimental realizations. For instance 
the spin values of $Fe II$ ions are $2$ and it is
experimentally found that these ions have anisotropy \cite{ref68}. Also for the 
disordered  $Fe_{(1-q)} Al_q$ alloys,
it has been   demonstrated that the effective spin value of iron  is $2$ and 
$5/2$ for Fe 2+ and Fe 3+ ions, respectively \cite{ref69}. Fe-Al alloys has been 
modeled and solved by EFT \cite{ref70}.   As in $S-3/2$ models, EFT is dominant 
method for the aim of determining critical and thermodynamical properties of the 
 $S-2$ Ising model. EFT is succesfully applied to the $S-2$ Ising model with 
transverse field \cite{ref71,ref72,ref73,ref74}, transverse crystal field 
\cite{ref75},  discrete distributed random field 
\cite{ref76,ref77}, random crystal field \cite{ref78},  site dilution problems 
\cite{ref79,ref80} and the model with biaxial crystal field \cite{ref81}. MFA 
has also been applied to the $S-2$ Ising model with transverse field 
\cite{ref82,ref83}  and random crystal field \cite{ref84}. $S-2$ Ising model on 
the Bethe lattice is the example of the $S-2$ Ising model with different geometry 
\cite{ref85,ref86}.

$S-5/2$ Ising model has many experimental realizations as mentioned in 
\cite{ref87}. This model has been solved within the  EFT \cite{ref88}. Also 
$S-5/2$ Ising model with transverse field \cite{ref89}, and quenched site dilution 
\cite{ref90} problems were solved with EFT. Some other methods exist  such as mean field 
renormalization group technique for the $S-5/2$ Ising model with random 
transverse field \cite{ref91}.

As seen from this short literature summary, many attempts have been made for the 
thermodynamical properties of the Spin-S model. But there is less attention paid 
on the hysteresis behaviors.  The aim of this work is to determine the hysteresis 
properties of the Spin-$S$ BC model and to obtain some general results especially 
about the multiple hysteresis behaviors. EFT for higher spin Ising model has 
been used in order to investigate hysteresis behaviors of the spin-S BC model. 
First attempts of the constructing formulation can be found in Ref.  
\cite{ref92} and decoupling approximation has been constructed in \cite{ref93}. The 
most advanced version of the formulation can be found in a review 
article \cite{ref94}.

Very recently, it has been shown by the author that, crystal field diluted $S-1$ 
BC model could exhibit double and triple hysteresis behaviors at large negative 
values of the crystal field and the physical mechanisms behind  these  
behaviors have been explained \cite{ref95}. Also it has been demonstrated that isotropic 
Heisenberg model 
could not exhibit these types of behaviors\cite{ref96}. 
The aim of the paper is obtain general results about the multiple hysteresis 
behaviors of the higher spin valued systems.  For this aim, the
paper is organized as follows: In Sec. \ref{model} we briefly
present the model and  formulation. The results and discussions are
presented in Sec. \ref{results}, and finally Sec. \ref{conclusion}
contains our conclusions.

\section{Model and Formulation}\label{model}

The Hamiltonian of the spin-S BC model with uniform longitudinal
magnetic field is given by
\eq{denk1}{\mathcal{H}=-J\summ{<i,j>}{}{s_is_j}-D\summ{i}{}{s_i^2}-H\summ{i}{}{
s_i},}
where $s_i$ is the $z$ component of the spin operator and it takes number of 
 $2S+1$ different values
such as $s_i=-S,-S+1,\ldots S-1,S$, $J>0$ is the ferromagnetic
exchange interaction between the nearest neighbor spins, $D$ is
the crystal field (single ion anisotropy), $H$ is the
external longitudinal magnetic field. The first summation in Eq.
\re{denk1} is over the nearest-neighbor pairs of spins and the other
summations are over all the  lattice sites.

We can construct the EFT equations by starting with generalized Callen-Suzuki 
\cite{ref97} identities, which are generalized versions of the identites for the 
$S-1/2$ system \cite{ref98,ref99} and given as

\eq{denk2}{\sandd{s_0^i}=\sandd{\frac{Tr_0 s_0^i
\exp{\paran{-\beta \mathcal{H}_0}}}{Tr_0\exp{\paran{-\beta
\mathcal{H}_0}}}},} where, $i=1,2,\ldots, 2S$ for the spin-S Ising system and 
$Tr_0$ is the partial trace over the site
$0$. Here, $\mathcal{H}_0$ denotes to the all interactions of the spin that 
belongs to the site $0$ and it has two parts as spin-spin interactions (denoted 
as $E_0$) and interactions with the fields (crystal field and magnetic field). 
From the Hamiltonian of the system represented  by Eq. \re{denk1},  $\mathcal{H}_0$ is 
given by 

\eq{denk3}{\mathcal{H}_0=s_0\paran{E_0+H}+s_0^2D, \quad 
E_0=\summ{\delta=1}{z}{s_\delta},
} where $z$ is the number of nearest neighbor interactions i.e. coordination number and 
$s_\delta$ is the nearest neighbor of the spin located at site $0$.  Inserting 
Eq. \re{denk3} into Eq. \re{denk2} and performing partial trace operations 
produces the equations in closed form as,
\eq{denk4}{\sandd{s_0^i}=\sandd{F_i\paran{E_0,H,D}}.
} By using differential operator technique \cite{ref100}, this equation can be 
written in the form

\eq{denk5}{
\sandd{s_0^i}=\sandd{\exp{\paran{E_0\nabla}}}F_i\paran{x,H,D},
} where $\nabla$ is the differential operator with respect to $x$ and the effect 
of the exponential differential operator on an arbitrary function 
$F(x)$ is given by
\eq{denk6}{\exp{\paran{a\nabla}}F\paran{x}=F\paran{x+a},} where $a$ is any 
constant. 
In this step, the exponential operator has $s_\delta$ terms in the exponent and for 
the aim of constructing the equations, we have to obtain polynomial forms of the 
equations in $s_\delta$. In order to get polynomial form of Eq. \re{denk5}, 
van der Waerden identities are often used: 
\eq{denk7}{
\exp{\paran{as_k}}=\summ{i=0}{2S}{A_i\paran{a}s_k^i}.
} The unknown coefficients $A_i\paran{a}$ can be obtained via the solutions of 
the $2S+1$ equations. These equations can be obtained by writing $2S+1$ possible 
values of the $s_k=-S,-S+1,\ldots, S-1,S$ in Eq. \re{denk7}. By this way the 
nonlinear equation system which includes number of $2S+1$ equations for the 
spin-S BC model can be obtained by Eq. \re{denk5} and solved numerically. But in 
order to avoid mathematical difficulties let us use approximated  van der 
Waerden identities which were proposed for the higher spin 
problems \cite{ref101} and it is given as
\eq{denk8}{
\exp{\paran{as_k}}=\cosh{\paran{a\eta}}+\frac{s_k}{\eta} \sinh{\paran{a\eta}},
} where $\eta^2=\sandd{s_k^2}$. Note that this is the first approximation made 
in the formulation, but as discussed in \cite{ref101} it produces accurate 
enough results in comparison with those obtained using the exact van der Waerden 
identity. Since this approximation equates $s_k^{2n}$ to $\sandd{s_k^2}^n$ and
$s_k^{2n+1}$ to $s_k\sandd{s_k^2}^n$, the number of $2S+1$ equations in Eq. 
\re{denk5} reduces to two. By using Eq. \re{denk8} these two equations can be 
written as

\eq{denk9}{m=\sandd{s_0}=\sandd{\prodd{k=1}{z}{\left[\cosh{\paran{J\eta\nabla}}
+\frac{s_k}{\eta} \sinh{\paran{J\eta\nabla}}\right]}}F_1\paran{x,H,D},
}
\eq{denk10}{\eta^2=\sandd{s_0^2}=\sandd{\prodd{k=1}{z}{\left[\cosh{\paran{
J\eta\nabla}}+\frac{s_k}{\eta} 
\sinh{\paran{J\eta\nabla}}\right]}}F_2\paran{x,H,D}.
}

Decoupling the terms $s_k$ and $s_k^2$ in the expanded form of Eqs. \re{denk9} 
and \re{denk10} and using the translationally invariance property of the 
lattice 
we arrive the equations:

\eq{denk11}{m=\left[\cosh{\paran{J\eta\nabla}}+\frac{m}{\eta} 
\sinh{\paran{J\eta\nabla}}\right]^zF_1\paran{x,H,D},
}
\eq{denk12}{\eta^2=\left[\cosh{\paran{J\eta\nabla}}+\frac{m}{\eta} 
\sinh{\paran{J\eta\nabla}}\right]^zF_2\paran{x,H,D}
.} The explicit forms of the functions in Eqs. \re{denk11} and \re{denk12} can 
be obtained by performing partial trace operations in Eq. \re{denk2} for 
arbitrary $S$ and they are given as
\eq{denk13}{F_1\paran{x,H,D}=\frac{\summ{k=-S}{S}{}k\exp{\paran{\beta D 
k^2}\sinh{\left[\beta k\paran{x+H}\right]}}}{\summ{k=-S}{S}{}\exp{\paran{\beta D 
k^2}\cosh{\left[\beta k\paran{x+H}\right]}}},
}

\eq{denk14}{F_2\paran{x,H,D}=\frac{\summ{k=-S}{S}{}k^2\exp{\paran{\beta D 
k^2}\cosh{\left[\beta k\paran{x+H}\right]}}}{\summ{k=-S}{S}{}\exp{\paran{\beta D 
k^2}\cosh{\left[\beta k\paran{x+H}\right]}}}.
}

By expanding Eqs. \re{denk11} and \re{denk12} with the help of Binomial 
expansion and applying Eq. \re{denk6} we get the equations as

\eq{denk15}{ m=\summ{n=0}{z}{}{}\komb{z}{n}\frac{m^n}{\eta^n}A_{n}^{(z)},
} 
\eq{denk16}{ \eta^2=\summ{n=0}{z}{}{}\komb{z}{n}\frac{m^n}{\eta^n}B_{n}^{(z)},
} and the coefficients are given by 

\eq{denk17}{
A_{n}^{(z)}=\frac{1}{2^z}\summ{p=0}{z-n}{}\summ{q=0}{n}{}\komb{z-n}{p}\komb{n}{q
}
(-1)^{q}
F_1[\eta J (z-2p-2q)], }
\eq{denk18}{
B_{n}^{(z)}=\frac{1}{2^z}\summ{p=0}{z-n}{}\summ{q=0}{n}{}\komb{z-n}{p}\komb{n}{q
}
(-1)^{q}
F_2[\eta J (z-2p-2q)]. }

By solving the system of nonlinear equations given by  Eqs.
\re{denk15} and \re{denk16}  with the coefficients given by Eqs.  \re{denk17} 
and \re{denk18}  we
get EFT - DA results for the spin-S BC model. Linearization  of Eqs. 
\re{denk15} and \re{denk16} in $m$ will yield equation system for the second order 
critical temperature.    Detailed investigation of the formulation used here can 
be found  in review article \cite{ref94}.


\section{Results and Discussion}\label{results}

For the numerical calculations, following  scaled (dimensionless) quantities have 
been used

\eq{denk19}{ d=D/J,t=k_BT/J,h=H/J. }

The hysteresis loops can be obtained for a given parameter set
($d,t$) by calculating the $m$ according to the procedure given
above, and by  sweeping the longitudinal magnetic field from $-h_0$
to $h_0$ and then in reverse direction (i.e. $h_0\rightarrow - h_0$). We
study on simple cubic lattice (i.e. $z=6$) within this work.

\subsection{Phase diagrams}

\begin{figure}[h]
\epsfig{file=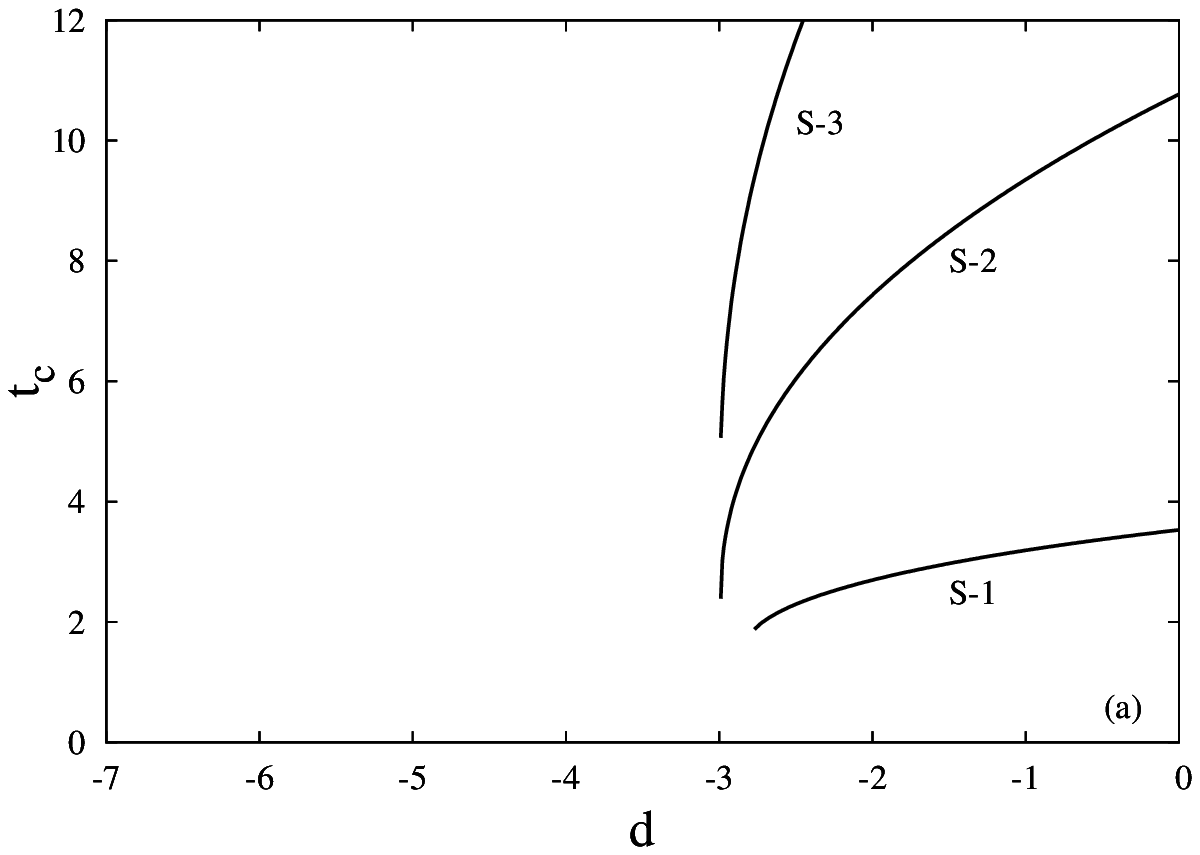, width=7cm}
\epsfig{file=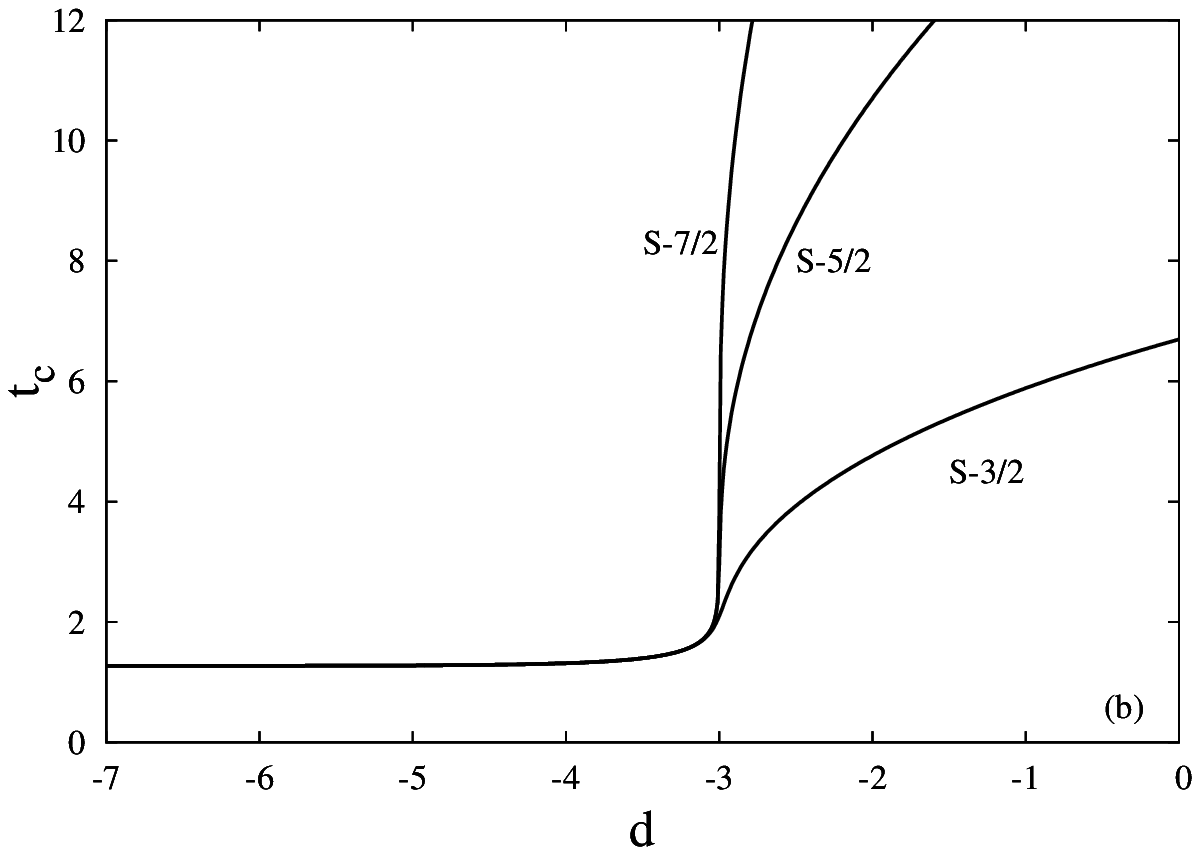, width=7cm}
\caption{Variation of the critical temperature of the spin-S BC 
model with crystal field. Fig. 1 (a)  contains integer spins while (b) contains 
half integer spin models.}\label{sek1}
\end{figure}

The phase diagrams of the spin-S BC model in $(d-t)$ plane can be seen in Figs. 
\ref{sek1} (a) for integer S and (b) for half integer S. Indeed these diagrams 
have already been obtained in the literature with several methods. The phase 
diagrams for the $S-3/2$ BC model were obtained within thermodynamically 
self-consistent Ornstein-Zernike approximation \cite{ref24}, MC \cite{ref25}, 
EFT based on differential operator technique \cite{ref19}, EFT  based on 
probability distribution technique \cite{ref21} and pair approximation 
\cite{ref23}. The phase diagrams for the $S-2$ BC model obtained within MFA 
\cite{ref83}, EFT \cite{ref71,ref73,ref76} and 
also for the $S-5/2$ BC model within EFT \cite{ref88}.

The main qualitative difference between the phase diagrams of integer (Fig. 
\ref{sek1} (a)) and half integer (Fig. \ref{sek1} (b)) spin model is seen for 
the  large negative values of the crystal field. Integer spin models have non 
magnetic ground state, while   
half integer spin models have magnetic ground state for the large negative 
values of $d$. This is because of the fact that, for the large negative values 
of the crystal field, only the minimum value of the $s_i$ makes the expectation 
value of the  Hamiltonian given in Eq. \re{denk1} minimum. This minimum value is 
$s=0$ for the integer S, and $s=\pm 1/2$ for half integer $S$. This means that for 
a ground state, rising  $d$ starting from  $d\rightarrow -\infty$ changes 
ground state from $s=0$ to $s=S$ and $s=1/2$ to $s=S$ for integer spin and half 
integer spin-$S$  BC model, respectively. These facts have been obtained by MFA 
\cite{ref6},  EFT \cite{ref11,ref47}, pair approximation \cite{ref23}, pair 
approximation and MC \cite{ref15}. It has been demonstrated in ground state 
that, for $S-3/2$ and $S-5/2$ BC models this transition occurs at  $D/J=-z/2$ 
where $z$ is the coordination number \cite{ref19,ref88}. 

Another qualitative difference between the phase diagrams of integer (Fig. 
\ref{sek1} (a)) and half integer (Fig. \ref{sek1} (b)) spin model is tricritical 
point. As seen  in Fig. \ref{sek1} (a) and (b), integer spin model has tricritical 
point while half integer spin model has not.

Note that, in Fig. \ref{sek1} (b) the phase diagram that lie along the large negative 
values of crystal field has temperature value $t_c=1.268$ which is just the 
critical temperature of the spin-1/2 model on simple cubic lattice.

\subsection{Hysteresis behaviors}

As demonstrated in Ref. \cite{ref95}, $S-1$ BC model has double hysteresis 
behavior for low temperature and large negative values of crystal field. This is 
due to the fact that, all spins 
of the system is in the state $s=0$ for low temperature and large negative 
values of crystal field. Magnetic field can induce transitions from this state 
to $s=1$ and $s=-1$ states. In general, as seen in Fig.  \ref{sek1}, Spin-$S$ BC 
model has ground state $s=0$ for integer S and $s=1/2$ for half integer S for 
large negative values of the crystal field. Then it is reasonable to think of 
that, the magnetic field can induce transition from $s=0$ to the state $s=1$ 
and further increasing field can induce transition from this state to $s=2$ and 
so on. In a similar manner,  for a half integer S, rising longitudinal magnetic 
field induce a transition from $s=1/2$ to the state $s=3/2$, if the magnetic field 
increases further this can cause transition to  $s=5/2$ and so on. This plateau 
behavior of the magnetization has been demonstrated for antiferromagnetic 
$S-3/2$ and $S-2$ systems \cite{ref30} and for $S-1$, $S-3/2$ and $S-5/2$ models 
on Bethe lattice \cite{ref36}. If this behavior is history dependent, then this 
means that the model could show multiple hysteresis behavior.

\begin{figure}[h]
\epsfig{file=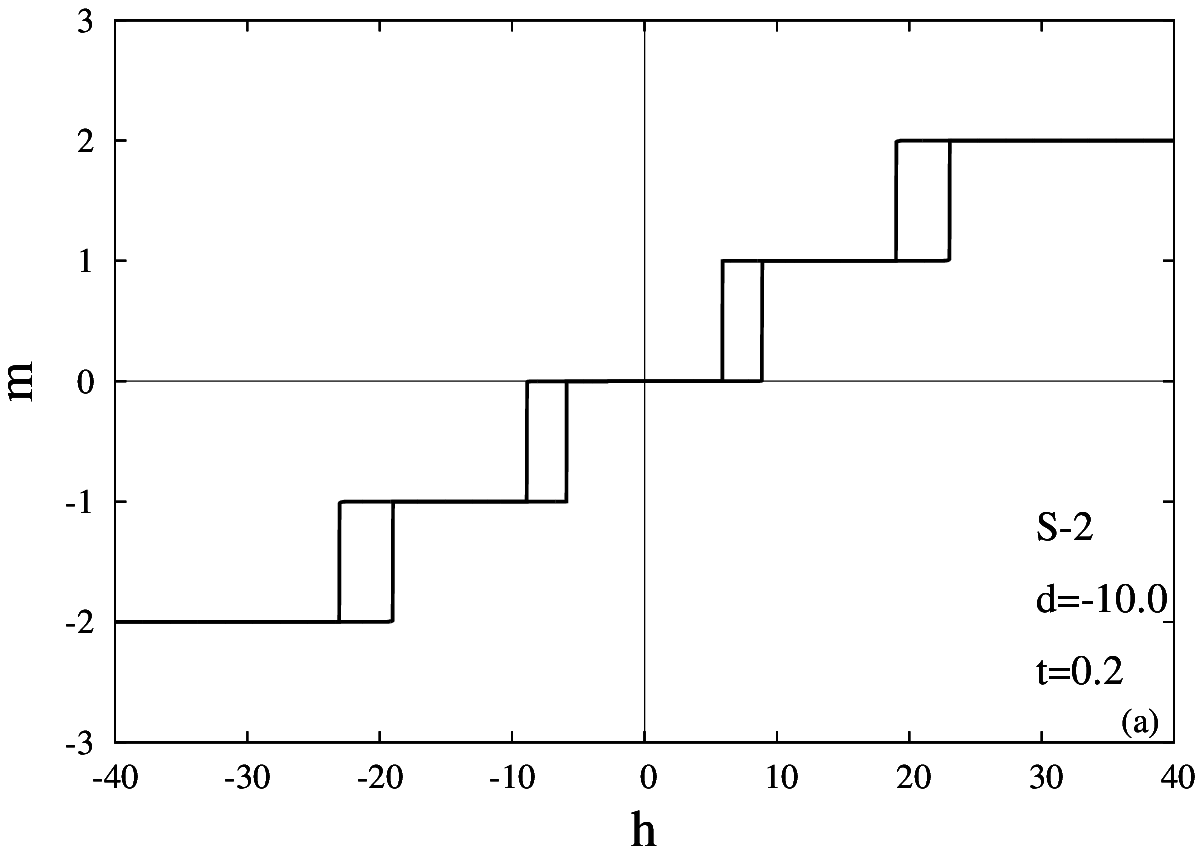, width=7cm}
\epsfig{file=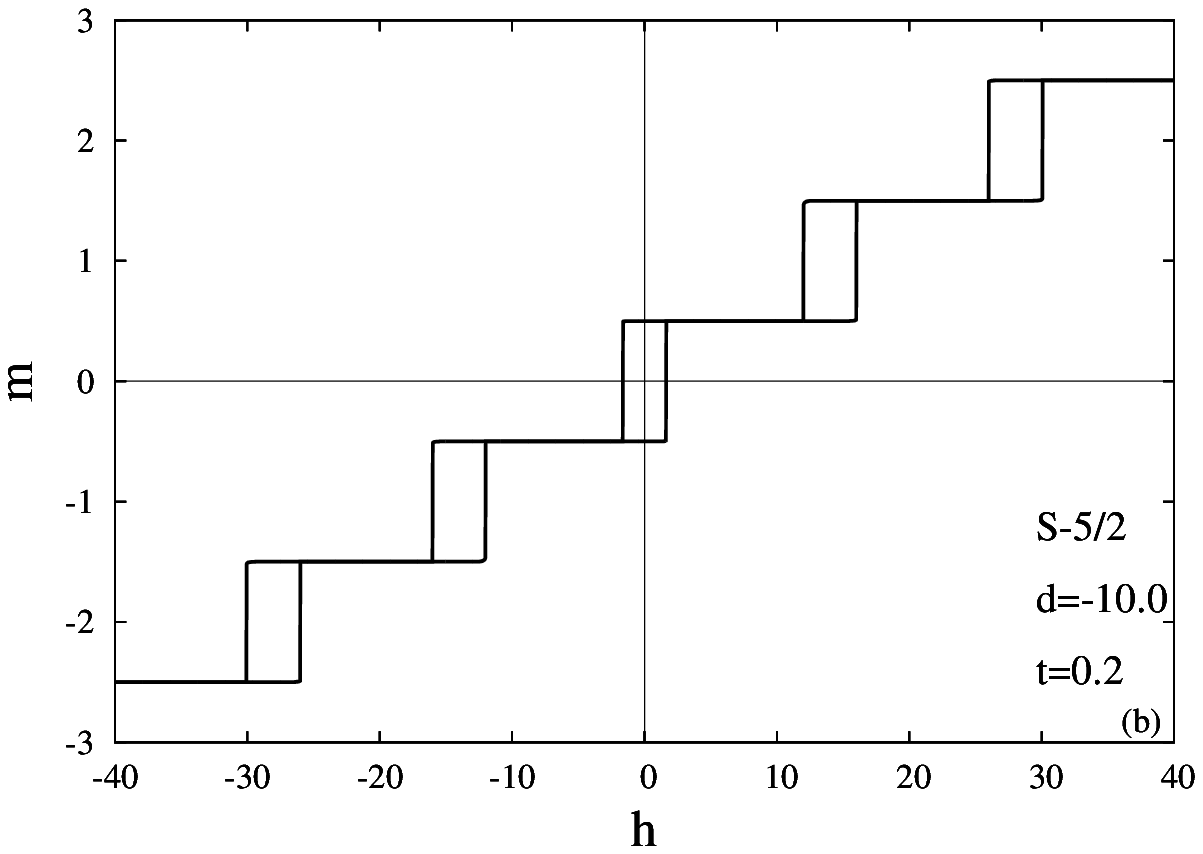, width=7cm}
\epsfig{file=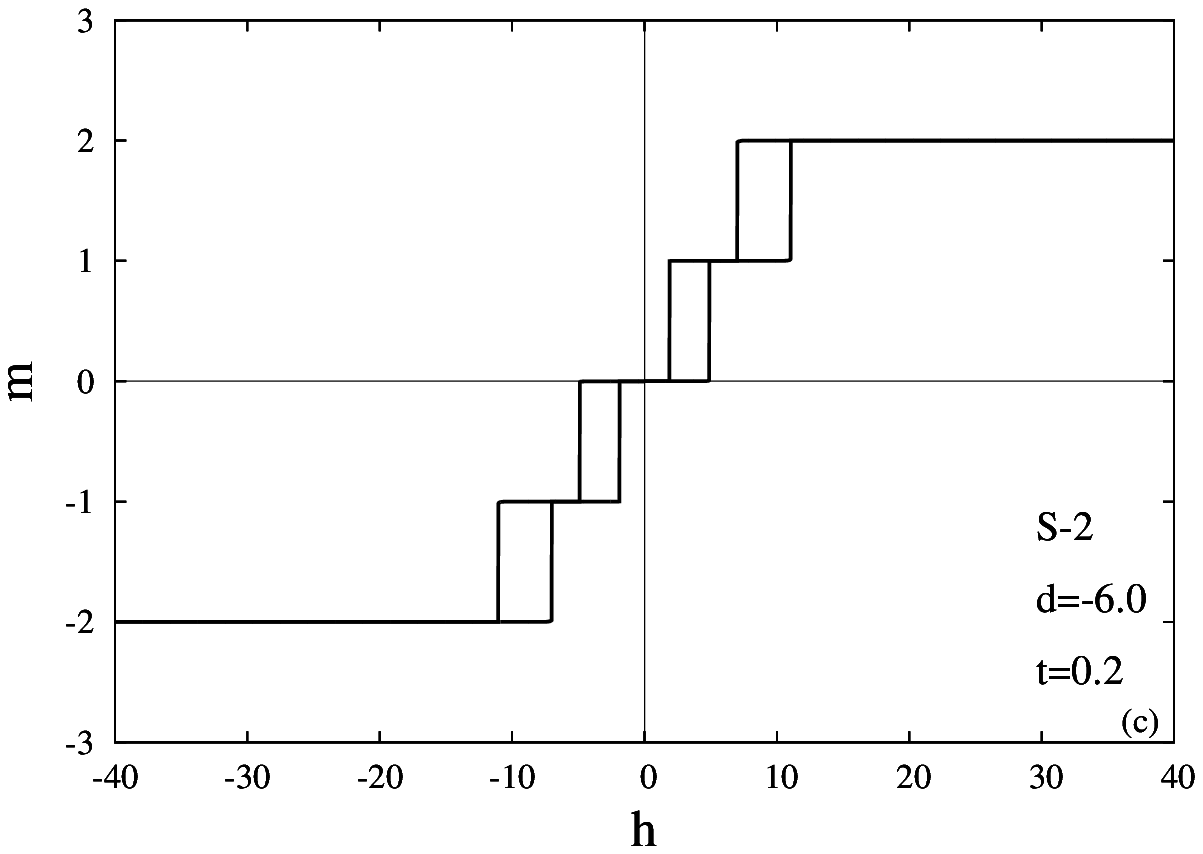, width=7cm}
\epsfig{file=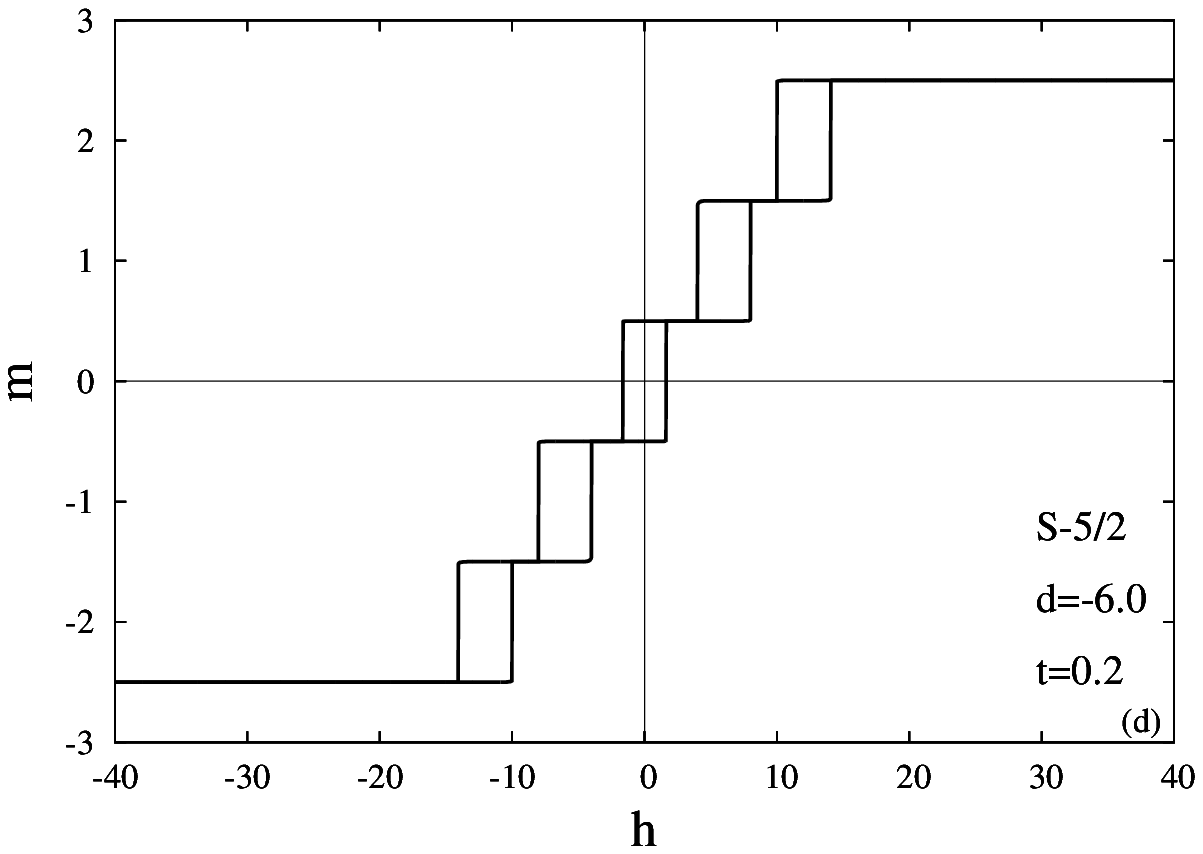, width=7cm}
\epsfig{file=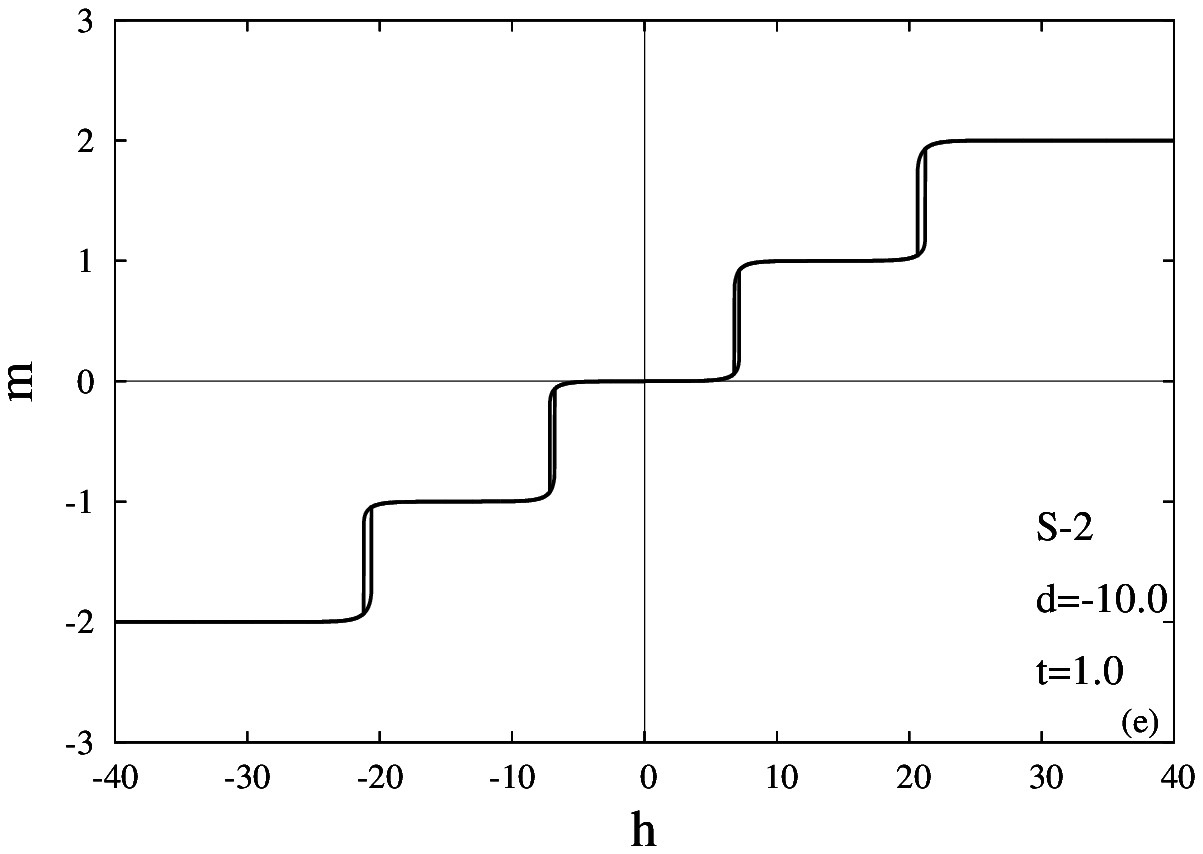, width=7cm}
\epsfig{file=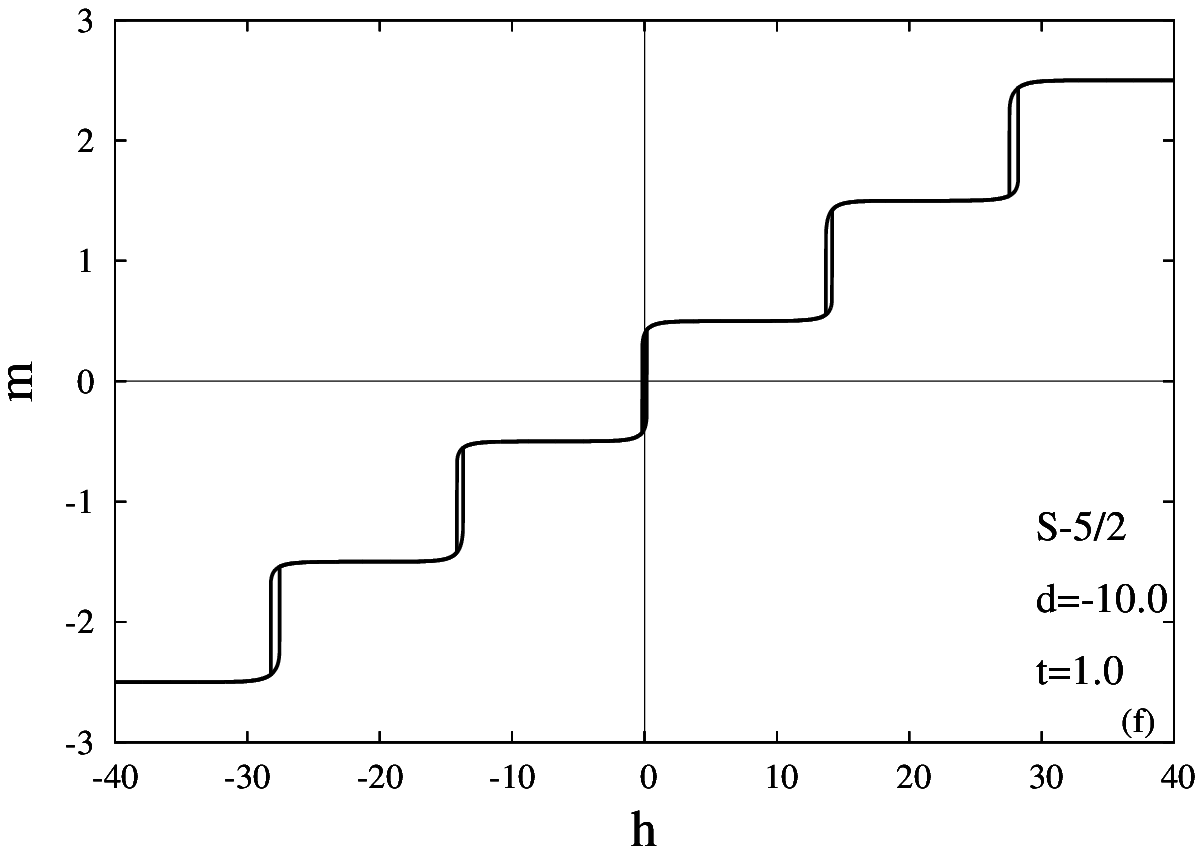, width=7cm}
\caption{Hysteresis loops of $S-2$ and $S-5/2$ BC models for various values 
of $d$ and $t$.}\label{sek2}
\end{figure}

In order to visualize these situations we depict some hysteresis loops of $S-2$ 
and $S-5/2$ BC models for selected values of Hamiltonian parameters and the 
temperature in Fig. \ref{sek2}.  Indeed spin-S BC model has hysteresis with 
$2S$-windows as seen in Figs. \ref{sek2} (a) and (b), for large negative values 
of $d$ and low temperature. The difference between the half integer and integer 
S model is that, the hysteresis loop of the half integer model has central 
window (which is symmetric about the origin) while the hysteresis loop of the 
integer model has not central loop (compare Figs. \ref{sek2} (a) and (b)). This 
is due to the fact that, half integer model has ordered phase in large negative 
values of crystal field and low temperature, while integer spin model has not. 
We can conclude by comparing Figs. \ref{sek2} (a) and (c), (b) and (d) with each other that, 
rising crystal field in negative direction causes  the windows to be aparted. 
Besides, the effect of the rising temperature can be seen by comparing Figs. 
\ref{sek2} (a) and (e), (b) and (f). Rising temperature causes a shrink of the 
windows.

\begin{figure}[h]
\epsfig{file=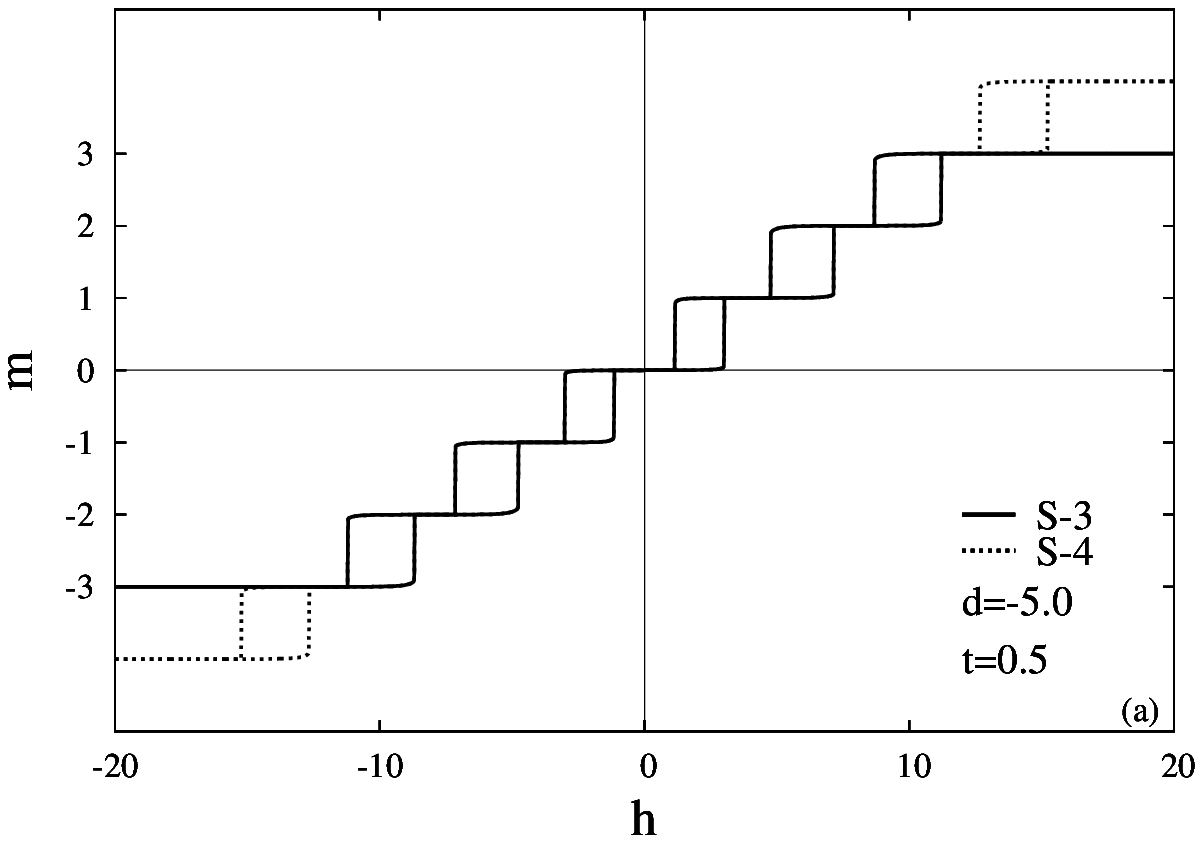, width=7cm}
\epsfig{file=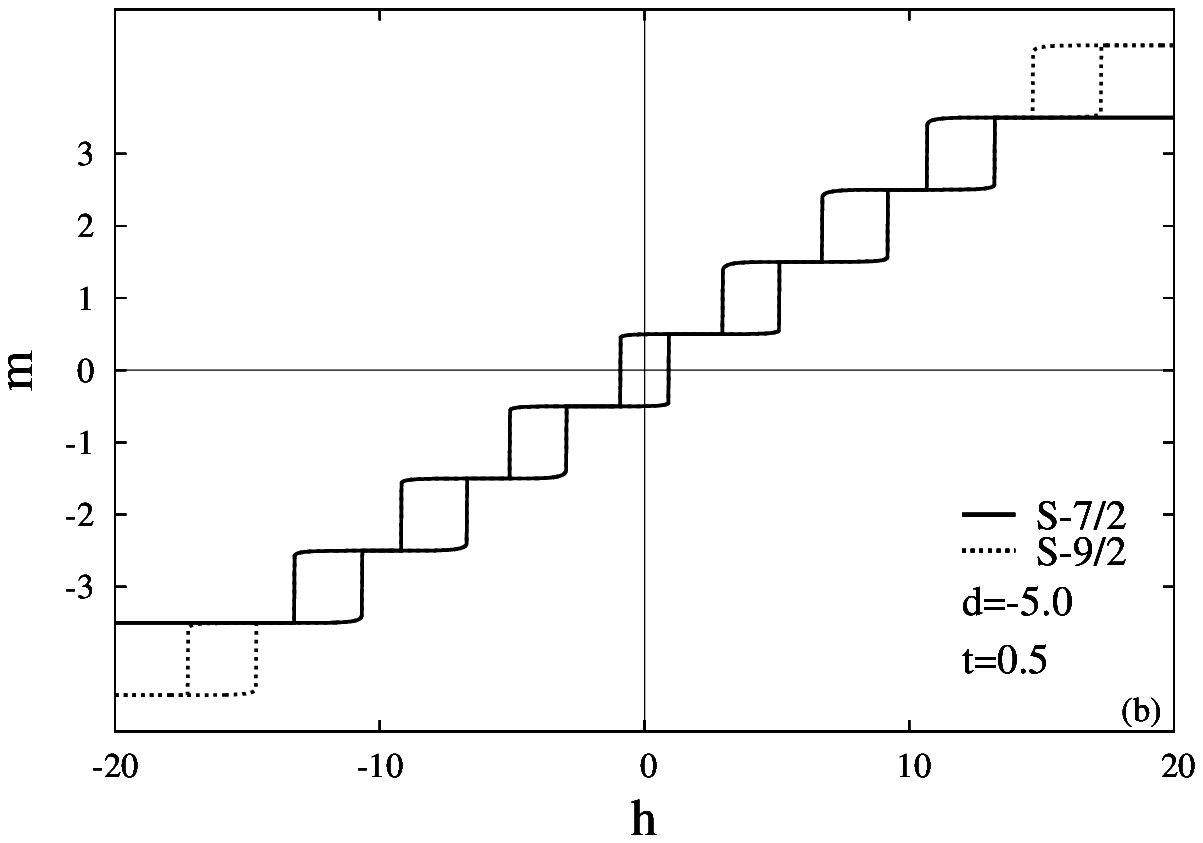, width=7cm}
\epsfig{file=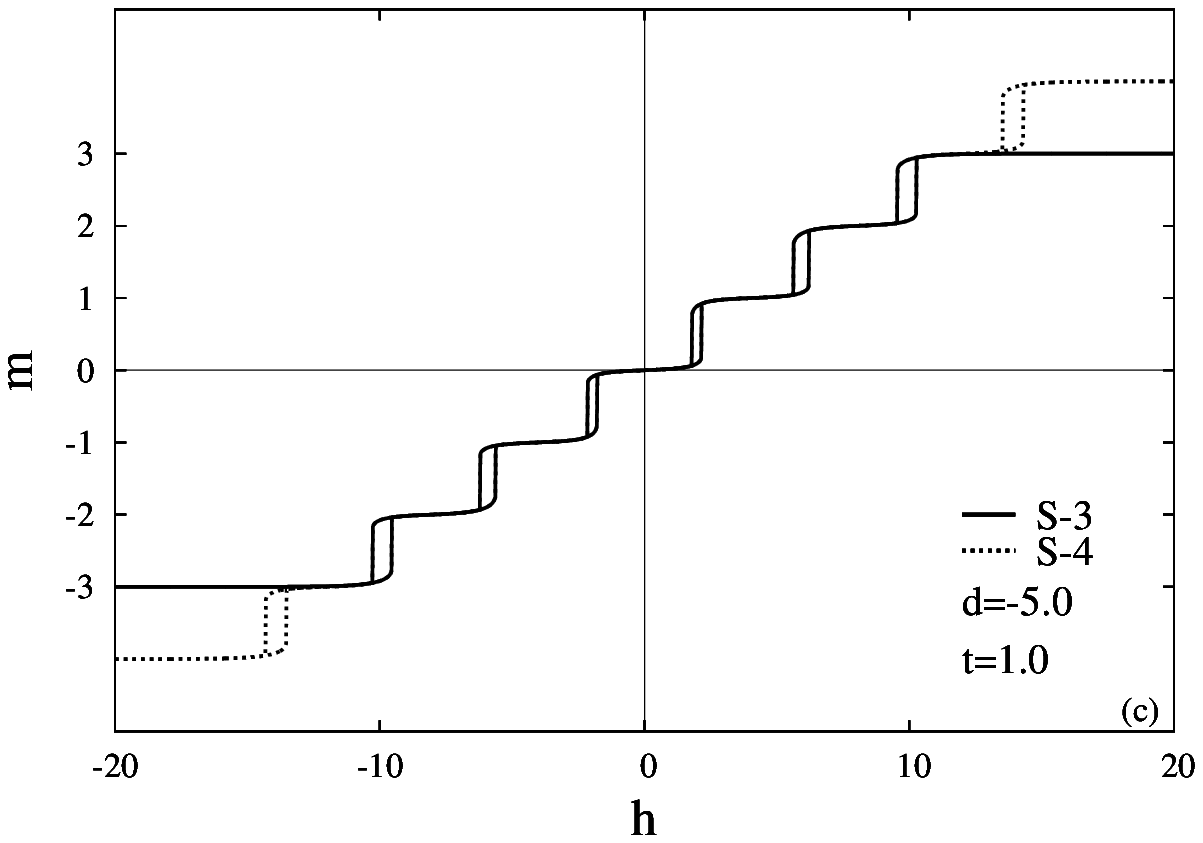, width=7cm}
\epsfig{file=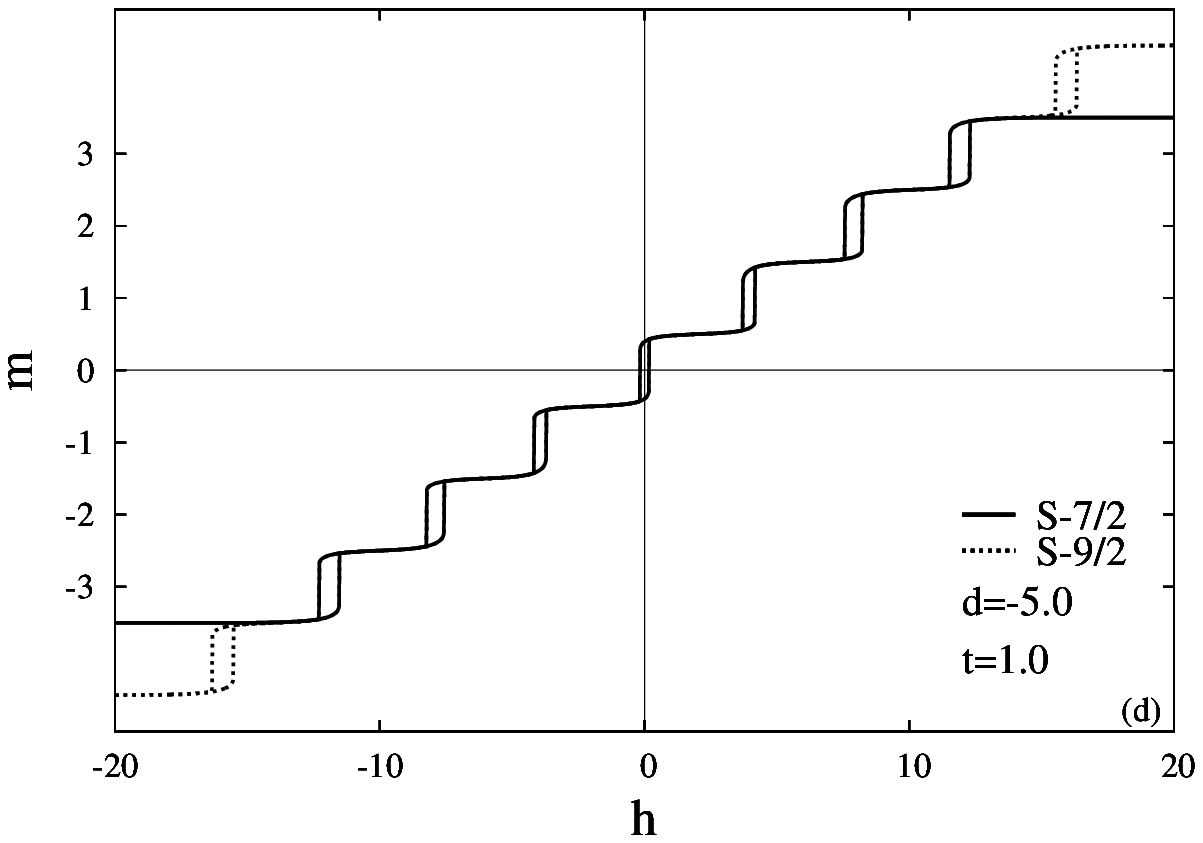, width=7cm}

\caption{Hysteresis loops of the S-3, S-4 and S-7/2, S-9/2 BC models for 
selected values of $d=-5.0$ and $t=0.5,1.0$.}\label{sek3}
\end{figure}

Interestingly, the hysteresis loop of spin-S and spin-(S+1) coincides, the only 
difference of latter hysteresis loop from the former loop is the outer windows 
(which are windows that appear in large negative and large positive values of 
the magnetic field). This can be seen in Fig. \ref{sek3}. Note that this nested 
structure of hysteresis loops of the successive Spin-S BC model is independent 
of Hamiltonian parameters and the temperature (compare Figs. \ref{sek3} (a) and 
(c), (b) and (d)). This structure can be interpreted as follows: transition from 
 $s=n$ to $s=n+1$ state requires the same amount of energy (which is supplied 
by the longitudinal magnetic field to the system in the case of hysteresis), 
regardless of the spin magnitude of the model. In other words the energy difference 
between $s=n$ and $s=n+1$ states is the same for all values of $S$.

Rising temperature causes shrink of the windows, as mentioned above. This 
shrinking behavior can be seen most dominantly on the closest windows to the center of 
the hysteresis loop, as temperature rises. This fact can be seen by comparing 
the central window of the hysteresis loops given in Figs. \ref{sek2} (b) and (f) 
or  \ref{sek3} (b) and (d). The central window almost disappears, while the 
other windows survive. This shrinking process continues with disappearing of 
windows starting from the closest ones to the center of the loop, with rising 
temperature. While the temperature rises, the number of windows of the hysteresis 
loops of the integer spin-S BC model changes as $2S\rightarrow 2S-2 \rightarrow 
\ldots\rightarrow 2\rightarrow 0$. Similar sequence is valid for the half integer 
spin-S BC model as $2S\rightarrow 2S-1 \rightarrow 2S-3 \rightarrow 
\ldots\rightarrow 2\rightarrow 0$. Hysteresis loop that has $0$ number of 
windows means that, there is no history dependent variation of the order 
parameter on the longitudinal magnetic field. Note also that, all loops in given 
sequences are the loops of the paramagnetic phase except $2S$ windows loop for the 
half integer spin-S BC model. The transition of this loop to the $2S-1$ windowed 
loop occurs at the critical temperature of the spin-S BC model, which is 
independent of the magnitude of the spin, $t_c=1.268$.

Let the temperatures $\tau^{i(S)}_n$ and $\tau^{h(S)}_n$ be the temperatures of 
transition from the $(n+2)$-windows loop to the $n$-windows loop of the spin-S 
BC model with integer and half integer S, respectively for large negative valued 
$d$. As seen in Fig. \ref{sek3}, rising temperature could not change the nested 
situation of the hysteresis loops of the spin-S and spin-(S+1) model. This means 
that, when the hysteresis loop of the spin-(S+1) model passes to the 
$(n+2)$-windows structure from the $n$-windows structure,  spin-S model passes 
to the $n$-windows structure from the $(n-2)$-windows structure. In other words, 
for the transition temperatures for large negative values of the $d$, following 
equalities hold,
$$
\tau^{i(S)}_n=\tau^{i(S+1)}_{n+2}, n=0,2,\ldots,2S-2
$$ 
$$
\tau^{h(S)}_n=\tau^{h(S+1)}_{n+2}, n=0,2,\ldots,2S-3
$$ 

\begin{figure}[h]
\epsfig{file=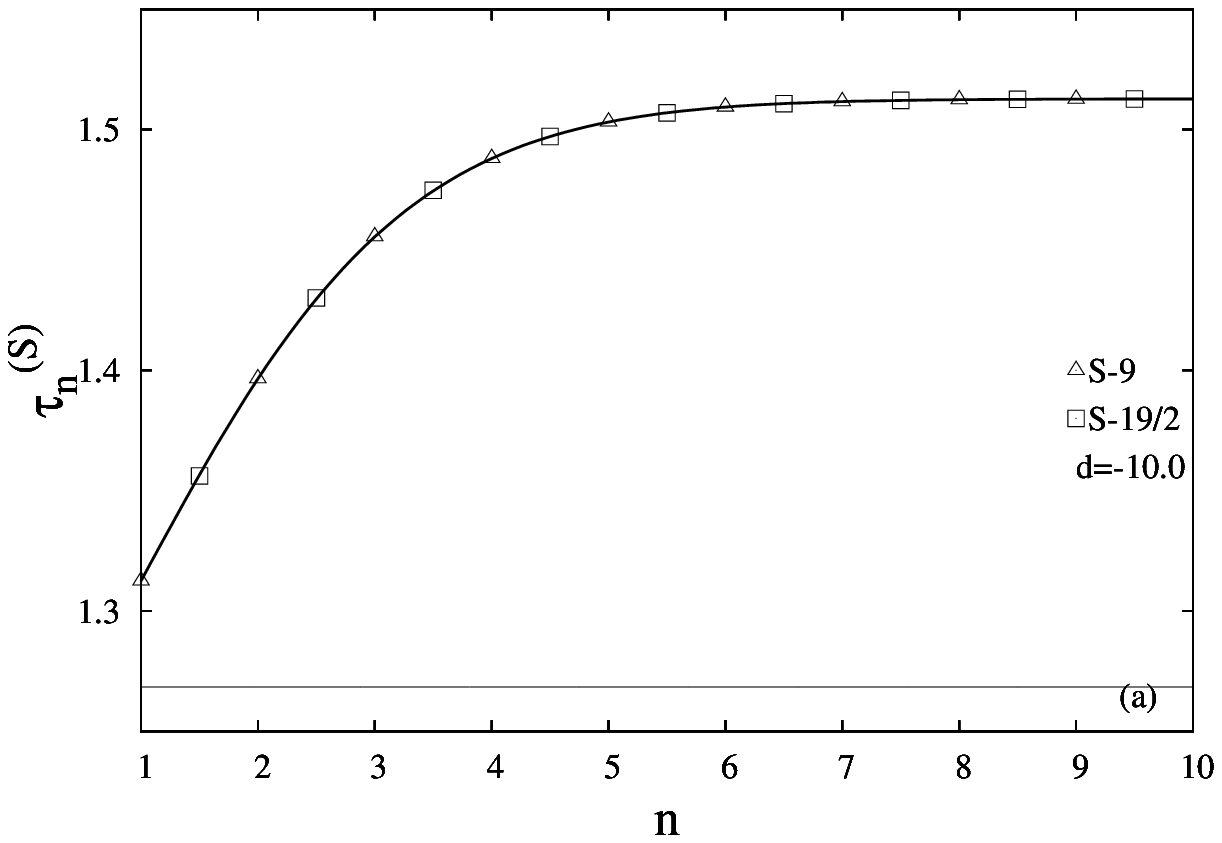, width=7.4cm}
\epsfig{file=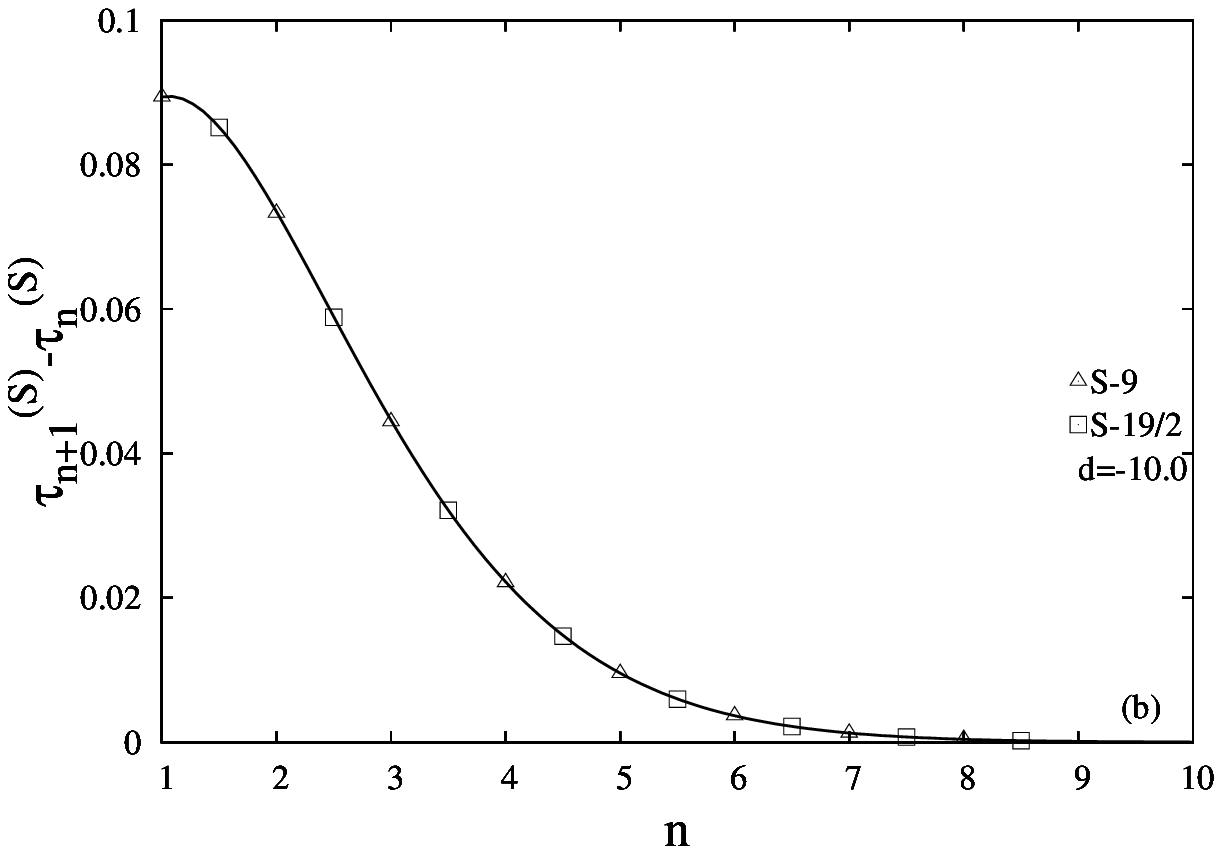, width=7.4cm}
\caption{The variation of (a) $\tau^{i(S)}_n$ and $\tau^{h(S)}_n$  and (b) 
$\tau^{i(S)}_{n+1}-\tau^{i(S)}_{n}$ and $\tau^{h(S)}_{n+1}-\tau^{h(S)}_n$, for 
$S-9$ (triangulars) and $S-19/2$ (squares) BC models. Lines are only guide for 
eye. Thin horizontal line represents the critical temperature of the half 
integer model for $d=-10.0$.}\label{sek4}
\end{figure}

The variation of both $\tau^{i(S)}_n$ and $\tau^{h(S)}_n$ (as well as 
$\tau^{i(S)}_{n+1}-\tau^{i(S)}_n$,  $\tau^{h(S)}_{n+1}-\tau^{h(S)}_n$ ) with $n$ 
can be seen in Fig. \ref{sek4} for $S-9$ and $S-19/2$ BC models. As seen in Fig. 
\ref{sek4} (a), both of the  transition temperatures of integer and half integer 
spin model lie on the same curve. Besides, the difference between the 
temperatures  of successive $n$ values decreases, as seen in Fig. \ref{sek4} 
(b).  This means as $S$ rising for spin-$S$ BC model, transitions between the 
higher $n$-windows loops could not be detected due to the low temperature 
differences.


\section{Conclusion}\label{conclusion}

Hysteresis characteristics of the general Spin-S ($S>1$) Blume-Capel model have 
been studied within the effective field
approximation. This work can be considered as the generalization of Ref. 
\cite{ref95}. In that work it has been demonstrated that, S-1 BC model has 
double hysteresis behavior for large negative values of the crystal field and 
low temperature. In general, spin-S BC model has $2S$-windowed hysteresis loop 
in the same region. The multiple hysteresis behavior is mostly attributed to 
different exchange interactions on some nanomaterials, such as nanowires. In 
this work simple physical mechanisms give rise to born of multiple hysteresis 
behavior given. Also the evolution of these multiple hysteresis loops with 
rising temperature and crystal  field have been given and discussed. Besides, some 
interesting results have been obtained such as nested characteristics of the hysteresis 
loops of successive spin-S Blume-Capel model.  

In Ref. \cite{ref96} it has been demonstrated that, the isotropic Heisenberg model could 
not exhibit multiple hysteresis behavior. Also crystal field dilution can induce 
transition from $2$ window hysteresis loop to the $3$ window loop in the S-1 Ising 
model. These results need to generalize and these generalizations will be our 
next works. We hope that the results
obtained in this work may be beneficial form both theoretical and
experimental points of view.

\end{document}